\documentclass[12pt, preprint]{aastex}

\shorttitle{Ultralight Energy}
\shortauthors{Nemiroff}

\begin{document}

\title{The Opposite of Dark Energy: Limits on $w = 2/3$ Ultralight Energy in the Early Universe}

\author{Robert J. Nemiroff}
\affil{Michigan Technological University, Department of Physics, \\
1400 Townsend Drive, Houghton, MI  49931}

\begin{abstract}
Might stable energy species ``lighter" than radiation, with $w > 1/3$, exist?  A dimensional expansion of the cosmological Friedmann Equation of energy has a clear place for them.  Such energies would affect the universe much differently than dark energies, and so are here dubbed ``ultralight."  As the universe expands, ultralight dilutes even faster than light.  Although any specie of energy can be mimicked by a properly evolving scalar field, ultralight energy species are hypothesized here to be stable and not related to dynamics of a scalar field.  Ultralight is not considered a candidate to make a significant contribution to the energy budget of the universe today, although ultralight might have affected the universe in the distant past.  In particular, the $w=2/3$ ultralight energy specie appears to have relatively mundane physical attributes. A discussion of properties and falsifiable attributes of ultralight is given.  The duration of primordial nucleosynthesis is extrapolated to limit the present density of $w=2/3$ ultralight to below one part in 100 billion of the critical density.
\end{abstract}


\maketitle

\section{Ultralight in a Standard Friedmann Cosmology}

Has humanity uncovered every possible form of energy?  Stable energy species can be classified by how they isotropically evolve in the cosmological Friedmann Equations.  As the universe scale factor $a$ expands, each energy specie will dilute as some power $n$ of $a^n$.  Well known examples include matter which dilutes as $a^3$, radiation which dilutes as $a^4$, and the cosmological constant which stays constant at $a^0$.   Currently unconfirmed energy species include cosmic strings which dilute as $a^2$ and domain walls which dilute as $a^1$.  In general, ``dark energy" indicates $n<3$, ``phantom energy" indicates $n<0$, and ``quintessence energy fields" indicate $0 \le n \le 1$.  So far, however, no published cosmological scenario has discussed cosmological limits on an energy species which dilute as $a^{>4}$. Such energy species will be called ``ultralight" here.  Could ultralight have existed in the early universe?

The idea of energy evolving with effective $w>1/3$ and $n>4$ has been hypothesized before in the context of time-varying scalar fields.  \citet{Kam90} discussed the possibility that an evolving scalar field could mimic an equation of state with $w>1/3$.  \citet{Spo93} and \citet{Joy98} discussed a scalar field evolving during an epoch with $w=1, n=6$, slowing the expansion rate of the universe as it exited inflation.  The ultralight species discussed here are not scalar fields and have no potential.  They are hypothesized to be stable energy forms that evolve in the standard way delineated by the established Friedmann Equations of isotropic General Relativistic cosmology.

In fact, ultralight energy species would fit quite naturally into the cosmological Friedmann Equations of General Relativity.  The Friedmann equation of energy is typically written \citep[see, for example,][]{Pea99} in the form 
 \begin{equation} \label{FriedmannOld}
 H^2 = {8 \pi G \over 3} \rho + {\Lambda \over 3}  - {k \over R^2} ,
 \end{equation}
where $H$ is the Hubble parameter, $G$ is the gravitational constant, $\rho$ is the energy density in matter, $\Lambda$ is the cosmological constant, $R$ is the scale factor of the universe, and $k$ is related to the curvature of the universe.   An expanded Friedmann equation can be written in dimensionless form as below: 
 \begin{eqnarray} \label{friedmann1}
 (H / H_o)^2    
 \sim           & \Omega_{\rm phantom \ energy} \ a^{<0} 
               + \Omega_{\rm cosmological \ constant} \ a^0  
               + \Omega_{\rm domain \ walls} \ a^{-1}  \nonumber \\
               + & \ \Omega_{\rm cosmic \ strings} \ a^{-2}  
               + \Omega_{\rm matter} \ a^{-3}  
               + \Omega_{\rm radiation} \ a^{-4}  
               + \Omega_{\rm ultralight} \ a^{>5}  \nonumber \\
               + & \ (1 - \Omega_{\rm total}) \ a^{2} , 
 \end{eqnarray}
where $a=R/R_o$ is the scale factor of the universe normalized to some epoch scale $R_o$, and each $\Omega_n$ above is fixed at the time when $a=1$.  In general here, it will be considered that $\Omega_{\rm total} = 1$ so that the trailing curvature term will be zero.

When any one energy specie dominates the energy density of the universe, the general solution to this Friedmann Equation of energy can be solved in a straightforward manner for scale factor $a$ as a function of time \citep[see, for example,][]{Pea99}.  Using $H = {\dot a}/a$, Friedmann's equation of energy can then be written 
 \begin{equation}
 \left( {1 \over a^2} \right) 
 \left( {da \over dt} \right)^2
 =  H_o^2 \ a^{-n} ,
 \end{equation}
where $H_o$ is Hubble's constant, the present value of the Hubble parameter.  When one energy specie with $n>0$ dominates, the solution to Friedmann's equation of energy becomes
 \begin{equation} \label{generalsolution}
 a = \left( { n \over 2} \right)^{2/n} \
       (H_o t)^{2/n} .
\end{equation}
In general, the higher the exponent $n$ for an energy specie, the more slowly a universe dominated by that energy specie will expand.  The above equation shows that an $n>4$ ultralight-dominated universe expands slowly compared to a $n \le 4$ non-ultralight dominated universe.  

\section{Physical Attributes of Ultralight}

Energy species are most simply defined as ``perfect fluids" involving only two parameters: pressure $P$ and density $\rho$.  When energy is conserved locally, $n$ is directly related to $w \equiv P/\rho$, the equation of state parameter, by the simple relation $n = 3(1+w)$ so that $w = n/3 -1$ \citep{Hut01}.  It is therefore clear that locally conserved ultralight not only has $n>4$ but $w>1/3$.   The term ``ultralight" energy is meant to contrast with ``dark" energy, such that dark energies are at one end of the energy specie spectrum with $n<3$ while ultralight is on the other end of the energy specie spectrum with $n>4$.  In analogy, ultralight is to light what ultraviolet light is to violet light.  

For a universe dominated by a single perfect-fluid energy specie where $w$ is unchanging, $c_s^2 = \partial P / \partial \rho$, where $c_s$ is the sound speed in the perfect fluid.  Therefore $c_s = \sqrt{w}$.  When $w < 0$ and $n < 3$ as for dark energy species, the sound speed $c_s$ becomes imaginary, causing instabilities.  Therefore, a frequent caveat imposed by cosmological analyses involving $w < 0$ energy species is to assume that the perfect fluid approximation is only valid on cosmological scales.  Such a caveat is not necessary for energy species $3 \le n < 6$ including the ultralight species with $4 < n \le 6$.  Ultralight species with $n>6$, however, will have a sound speed formally greater than the speed of light. 

A particularly mundane specie of ultralight energy has $n=5, w = 2/3$.   Previously, all of observationally detected energy species, $n=3$ matter, $n=4$ radiation, and $n=0$ cosmological constant, all dilute with universe expansion as integer powers of $a$.  The $n=5$ ultralight would dilute cosmologically as $a^5$, also an integer power of the universe scale factor $a$. Furthermore, $n=5$ ultralight can be a perfect fluid as its sound speed $c_s$ is neither imaginary nor greater than $c$.  Therefore, the $n=5$ ultralight energy specie will be highlighted here as particularly interesting.  When $n=5$ ultralight dominates the Friedmann Equations, Eq. (\ref{generalsolution}) shows that $a \sim t^{2/5}$.

\section{Falsifiability of Ultralight}

Why consider ultralight?  There is no evidence that ultralight exists in the present universe.  There is no present evidence that ultralight ever existed in the universe.  Ultralight is neither required nor expected by any theory of particle physics of which the author is aware.  
One reason to consider ultralight now, however, is because the discovery of dark energy has opened the door on the consideration of new energy species, including phantom energy \citep{Cal03} and many different types of quintessence \citep[see, for example,][]{Zla99}.  Ultralight is neither of those.  Also, ultralight has never been explicitly looked for.  This does not mean, however, that it does not exist, has never existed, and could not be found if looked for. Ultralight has well defined properties that are falsifiable -- specific searches could prove it exists definitively. 

Ultralight could be detected were the expansion history of the universe found to contain an epoch that lasted longer than could be explained by the domination of darker energy species.  A limiting example, given below, is the nucleosynthesis epoch, where the universe expansion time scale can be measured independently against nuclear decay time scales.

Where might ultralight have come from?  It is possible that very early in the universe, phase transitions occurred that created ultralight.  Following such a transition, ultralight might have been the dominant energy specie(s).  Since ultralight dilutes with universe expansion faster than other energy species, any ultralight epoch might have ended as the relative abundance of lower $n$ energy species increased.  Possibly, that ultralight might have diluted to a nearly imperceptible level by today.  

Why would any energy form act with a $n>4$?  Although there are clear dimensional paths to understanding most other integer $n$ species, there is no such clear path for understanding ultralight.  Possibly, ultralight might respond, at least partly, to other spatial dimensions, such as those spatial dimensions hypothesized in string theory\citep{Wit95} or Randall-Sundstrum cosmologies.\citep{Ran99}  If so, such energy species might only act as lower $n$ energy species as these dimensions became cosmologically unimportant.  Alternatively, ultralight might somehow incorporate multiple sensitivities to the time dimension.  

\section{Astrophysical Limits on Ultralight}

A limit on the past and present abundance of ultralight can be found from primordial nucleosynthesis results. Cosmological nucleosynthesis occurred at a temperature of about $10^9$ K and a redshift of about $z \sim$ 3 x 10$^8$ \citep{Pea99}. Previously, it has generally been presumed that radiation dominated the universe during nucleosynthesis, but we now consider the strange possibility that the universe was dominated instead by ultralight.  In that case, although density of $n=3$ baryonic nuclei might be unaffected, the time of nucleosynthesis onset and duration would be lengthened.    

Relatively strong limits can be set on the $n=5$ ultralight energy specie during primordial nucleosynthesis.  Then the expansion rate of the universe would have been $a \sim t^{2/5}$ instead of $a \sim t^{1/2}$ when dominated by $n=4$ radiation.  (An interesting caveat is that \citet{Bar82} proposed an anisotropic cosmology that could also slow the universe expansion rate and could therefore mimic the effects of a stable ultralight energy specie.)  Assuming nucleosynthesis stops at the same temperature $T$ and scale factor $a$, then the time that nucleosynthesis ends in a $n=5$ ultralight dominated universe $t_5$ would be related to the time in the $n=4$ radiation dominated universe $t_4$ by $t_5 \sim (2/5) \ 2^{5/4} \ t_4^{5/4}$.  Given that $t_4 \sim$ 180 sec, then $t_5 \sim$ 625 sec.  

The result of a longer nucleosynthesis epoch would be that more neutrons would decay leading to a lesser abundance of helium that predicted by radiation-dominated nucleosynthesis \citep{Oli00, Alv01, Nak06}.  Nevertheless, just like the limits on light elements in standard Big Bang Nucleosynthesis are used to constrain the number of neutrino species \citep{Oli00} by limiting the expansion rate of the universe, those limits can also be used to indicate that $n=5$ ultralight did {\it not} dominate the universe during nucleosynthesis, and had at most a density comparable to $n=4$ radiation.  It has been assumed here that ultralight remains stable and does not strongly interact with other energy species.  The corresponding present limit on the density of $n=5$ ultralight would therefore be a factor of $(1+z)^{5/4}$ lower than the present density of background radiation today.  Given that $\Omega_{\rm radiation} \sim 2$ x 10$^{-5}$ \citep{Pea99}, $\Omega_{\rm ultralight} < 10^{-11}$ today.

A limit on the past and present abundance of ultralight can also be found from cosmic microwave background radiation (CMBR) results. The microwave background results from the recombination era, which occurred at a temperature of 3000 K and at a redshift of about $z \sim 1100$ \citep{Pea99}.  It is generally presumed that $n=3$ matter dominated the expansion of the universe at recombination.  Introducing a significant density of ultralight at recombination, coupled with a flat universe, must result in a significantly reduction in the density of $n=3$ dark matter.  Such a reduction would be seen in the CMBR as a decrease in the scale of the temperature fluctuations. From analysis of microwave background data including recent data from the WMAP mission \citep{Spe03}, many cosmological parameters have been well determined, including the density of baryons at recombination \citep{Hu02}.  Therefore, conservatively, ultralight density must be lower than matter density at recombination.

This general density limit can be turned into more specific density limit on $n=5$ ultralight today.  Given that the present density of $n=3$ matter found from analyses of CMBR is about $\Omega_{\rm matter} \sim 0.25$, the present density of ultralight would then be constrained to be a factor of $(1+z)^{5/3}$ smaller, or $\Omega_{\rm ultralight} < 10^{-6}$.  The limit on $n=5$ ultralight found from nucleosynthesis turns out to be more confining.

Is it possible, though, that the missing energy between $\Omega_{\rm matter}$ and $\Omega_{\rm total}=1$ is not dark energy but really ultralight?  No.  Were the universe presently dominated by ultralight, it would be slowing down and not accelerating, an attribute that is significant discord with concordance cosmology that includes data from the dimness of distant supernovae \citep{Rie98, Per99}, the spot size distribution on the cosmic microwave background \citep{Spe03}, and the observed distribution of nearby galaxies \citep{Haw03}.  It is possible, though, that an ultralight epoch preceded the radiation epoch, and it was the ultralight epoch that started when inflation ended.  In this scenario, the radiation epoch started only when ultralight had diluted, through universe expansion, to a density comparable to radiation.

In sum, a new energy species dubbed ultralight has been hypothesized that evolves with universe scale factor differently than all other types of energy species.  In particular $n=5$ ultralight has the tame physical properties of expanding as an integer power of universe scale factor and having a positive sound speed less than $c$.  Nucleosynthesis results limit $n=5$ ultralight from being more than $\Omega_{\rm ultralight} \sim 10^{-11}$ today, although epochs dominated by ultralight energy species might have existed in the early universe.

\end{document}